# Energy currents in the field formed by waves with different frequencies

**Igor Mokhun\*, Igor Bodyanchuk, Kateryna Galushko, Yuriy Galushko, Oleksandr Val, Yuliia Viktorovskaya**

*Chernivtsi university, Kotsybinsky Str., 2, Chernivtsy 12, 58012, Ukraine*
*Corresponding author: i.mokhun@chnu.edu.ua*

**The behavior of instantaneous and averaged vectors of the Poynting vector transverse component for the resulting field formed as a superposition of waves with different frequencies and different polarizations is considered. Results of computer simulation and experimental data are obtained for the superposition of frequently differed vortices with different and the same signs of topological charges. It is shown that the captured particle rotates due to the orbital momentum and substantial increase of particle rotation speed in the case of polychromatic radiation *as opposed* to monochromatic is observed.**

http://dx.doi.org/10.1364/OL.99.099999

In recent years, a relatively new trend in modern photonics, singular optics [1-4], has been developing successfully. In the framework of this direction, the features of electromagnetic fields as the singularities of various types, their networks, and the physical manifestations of the special behavior of the wave in their vicinity are considered.

In this sense, particularly promising research is related to the fact that in the vicinity of optical singularity a wave acquires unique physical properties that are unusual for other optical structures. In the singularity region, a specific angular momentum of a field arises [5–8], energy fluxes are of a unique nature [9].

At the same time, the vast majority of the results were obtained in the framework of the coherent or monochromatic approximation. Only recently there has been an interest in the study of incoherent, polychromatic optical fields [10–16]. Only the first steps have been taken in this direction, but, in our opinion, it is here that significant progress should be expected in research on singular optics. This is especially true for the development of new rapidly developing technology of optical tweezers and manipulators [17-19].

This article presents our results in this direction devoted to some aspects of the formation of energy flows arising in polychromatic fields and fields formed as a result of different-frequency interference. All considerations will be carried out for the paraxial approximation.

In the study of energy flows, the behavior of the Poynting vector, we will analyze both averaged and instantaneous characteristics, since:

• if the coherence time is longer than the response time of the physical system, then instantaneous flows can play a significant role in the physical response of the system that is affected by the optical wave.

• different-frequency components can be interconnected, for example, via optical second harmonic generation (SHG) [14], effects of nonlinear optics [20], in ring resonators [21,22] or when using independent sources, lasers with a long length coherence, which can reach several kilometers [23], etc.

Finally, an understanding of how instantaneous flows are formed makes it possible to more clearly understand the mechanisms of formation of the averaged parameters of a wave.

**The instantaneous Poynting vector at a superposition of two waves with different frequencies**

It can be shown that in the case of the paraxial approximation, the transverse components of the instantaneous Poynting vector for superposition of two waves with different frequencies can be described as the following:

$$\begin{cases} P_x = P_{x1} + P_{x2} + In_x \\ P_y = P_{y1} + P_{y2} + In_y, \\ P_z = P_{z1} + P_{z2} + In_z \end{cases} \quad (1)$$

where $P_{li}$ – "coherent" Poynting vectors [3] corresponding to the components of each wave.

It is essential that, in (1), the so-called intermodulation components are included.

$$\begin{cases} In_x = -\frac{c}{4\pi}[\frac{1}{k_1}(E_{x2}T_{21} - E_{y2}T_{11}) + \frac{1}{k_2}(E_{x1}T_{22} - E_{y1}T_{12})] \\ In_y = -\frac{c}{4\pi}[\frac{1}{k_1}(E_{y2}T_{21} + E_{x2}T_{11}) + \frac{1}{k_2}(E_{y1}T_{22} + E_{x1}T_{12})], \\ In_z = \frac{c}{2\pi}(E_{x1}E_{x2} + E_{y1}E_{y2}) \end{cases} \quad (2)$$

where

$$\begin{cases} T_{1i} = E_{xi}\Phi_{xi}^{y} - E_{yi}\Phi_{yi}^{x} + \frac{a_{xi}^{y}}{a_{xi}}E_{xi,\pi/2} - \frac{a_{yi}^{x}}{a_{yi}}E_{yi,\pi/2} \\ T_{2i} = E_{xi}\Phi_{xi}^{x} + E_{yi}\Phi_{yi}^{y} + \frac{a_{xi}^{x}}{a_{xi}}E_{xi,\pi/2} + \frac{a_{yi}^{y}}{a_{yi}}E_{yi,\pi/2} \end{cases}, \quad (3)$$

$a_{li}$ and $\Phi_{li}$ are amplitude modulus and space phases of components, $a_{li}^{v}$ and $\Phi_{li}^{v}$ are partial derivatives from this values, $i=1,2$ – indices of the first and second wave, $k_i = \frac{2\pi}{\lambda_i} = \frac{\omega_i}{c}$ – wave numbers, $E_{li}, E_{li,\pi/2}$ – electric field strengths and shifted on $\pi/2$ ones correspondingly:

$$E_{li} = a_{li}\cos(\omega_i t + \Phi_{li} + k_i z), \quad (4)$$

$$E_{li} = a_{li}\sin(\omega_i t + \Phi_{li} + k_i z), \quad (5)$$

Naturally, the intermodulation components present in (1), in addition to the coherent components $P_{li}$, radically change the structure of the instantaneous flows of the resulting field.

## Averaged by time components of Poynting vector

Note that the "beat time" of a wave formed by two waves with different frequencies, which in the general case corresponds to the "complete closure" of the Lissajous trajectory of the electric vector [12,14,24] can be represented as:

$$T_b = \frac{\lambda^2}{c\Delta\lambda} + \frac{\lambda}{c}. \quad (6)$$

where $\lambda$ is the shorter wavelength, $\Delta\lambda$ is the difference between the wavelengths involved in the superposition.

The first term in this relation can be interpreted as the formally introduced coherence time [23]. Note that for small $\Delta\lambda$, the beat time differs little from the coherence time.

It is enough to simply show that when averaging over time significantly exceeds time (6), relations (1) are transformed to the form:

$$\begin{cases} \bar{P}_x = \bar{P}_{x1} + \bar{P}_{x2} \\ \bar{P}_y = \bar{P}_{y1} + \bar{P}_{y2}, \\ \bar{P}_z = \bar{P}_{z1} + \bar{P}_{z2} \end{cases} \quad (7)$$

where $\bar{P}_{xi}$ – "coherent" components of Poynting vectors associated with both waves.

In other words, we have the classical case of "incoherent" addition of quantities. Obviously, this conclusion is valid for any number of waves participating in a superposition. Moreover, nothing will change significantly when instead of a discrete set of waves we have a continuous spectrum, i.e. polychromatic wave. It can be shown that relation (8) transforms into integrals:

$$\bar{P}_l = \int_0^\infty \bar{P}_l(\lambda)d\lambda, \quad (8)$$

where $l = x, y, z$, and $\bar{P}_l(\lambda)$ - are the "spectral Poynting vector components".

Correspondingly the resulting angular momentum of the field is a simple sum of elementary moments created by individual waves.

$$M = M_1 + M_2. \quad (9)$$

Or for a polychromatic field, the resulting moment is an integral of the moments created by the "spectral" components:

$$M = \int_0^\infty M(\omega)d\omega, \quad (10)$$

The obtained relations were verified by computer simulation and experimental studies using the superposition of vortex beams with different frequencies as an example.

## Superposition of vortex beams

For simplicity, we considered the case when vortex beams of the same amplitude propagated coaxially and these vortices were linearly polarized in one plane. Thus, elementary fields can be considered in the scalar approximation.

For the scalar wave in contrast to the vector one [2], only longitudinal disclinations (instant field zeroes) are observed [3]. In other words, the averaged behavior of the Poynting vector is formed as the averaging of differently oriented "edge disclinations like" for each moment.

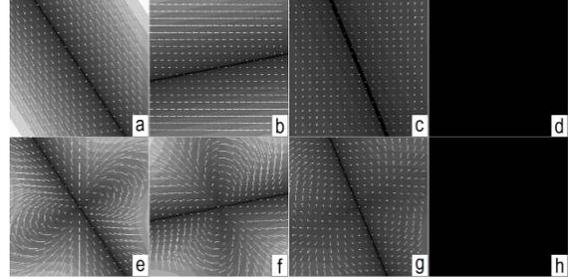

Fig. 1. The instantaneous Poynting vector for the resulting field, formed as a superposition of vortices with different frequencies and the same topological charges (*a-d*), different charges (*e-h*). The time difference between implementations is equal to 1/8 vibration period of the wave with a bigger wavelength.

We carried out a computer simulation of instantaneous energy flows for the superposition of vortex beams for both cases identical and different sign of topological charges.

Figure 1 (a-d) shows the behavior of the instantaneous component of the Poynting vector transverse component for 1/2 period of wave oscillations with greater $\lambda$ for the case when the vortices have the same signs of topological charges. As can be seen, the instantaneous components are realized as a system of rotating edge disclinations, similar to disclinations that arise in the scalar case [3]. However, the rotational speed of disclinations is two times higher and, in contrast to the scalar case, the total value of the component changes in time, up to complete attenuation.

Figures 1 (e-h) correspond for different signs of topological charges. As in the previous case, the instantaneous component is realized as a system of rotating edge disclinations. The rotation velocity of disclinations is also two times higher than in the scalar case. The magnitude of the component again changes in time, up to complete attenuation. As for the orientation of the instantaneous vector, it has a more complex character than in the previous case.

For the scalar case the components of the averaged Poynting vector of each elementary wave have the form [3]:

$$\bar{P}_t = \begin{cases} \bar{P}_x = -\frac{c\lambda}{16\pi^2}a^2\Phi^x \\ \bar{P}_y = -\frac{c\lambda}{16\pi^2}a^2\Phi^{y\prime} \end{cases} \quad (11)$$

where $a, \Phi^l, l = x, y$ are the amplitude and derivatives of the phase of the beam.

Let us consider isotropic vortices [3], "planted" on a Gaussian beam (Lager-Gaussian beam, Lager-Gaussian mode), for which the complex amplitude can be represented as:

$$U = \alpha\sqrt{x^2 + y^2}\exp(-\frac{x^2+y^2}{2\sigma^2})\exp[jS\arctan\left(\frac{y}{x}\right)], \quad (12)$$

where $\alpha$ – the amplitude coefficient, $\sigma, S$ – the beam width and the topological charge of the vortex, respectively

Then, the components of the Poynting vector have the form:

$$\bar{P}_t = \begin{cases} \bar{P}_x = \alpha S \frac{c\lambda}{16\pi^2} y\exp(-\frac{x^2+y^2}{2\sigma^2}) \\ \bar{P}_y = -\alpha S \frac{c\lambda}{16\pi^2} x\exp(-\frac{x^2+y^2}{2\sigma^2}) \end{cases}, \quad (13)$$

The expressions for the Poynting vector components for vortex beams with different frequencies differ only by the factor $\lambda$.

Due to the superposition of two waves with different frequencies, the resulting averaged vector is a sum of vectors, for two optical vortices with wavelengths $\lambda_1$ and $\lambda_2$ that differ by intensity to the coefficient $\gamma$ we have:

$$\bar{P}_t = \begin{cases} \bar{P}_x = \frac{c}{16\pi^2} y\exp(-\frac{x^2+y^2}{2\sigma^2})(\lambda_1 S_1 + \gamma\lambda_2 S_2) \\ \bar{P}_y = -\frac{c}{16\pi^2} x\exp(-\frac{x^2+y^2}{2\sigma^2})(\lambda_1 S_1 + \gamma\lambda_2 S_2) \end{cases}. \quad (14)$$

The ratio of the Poynting vector values for the case when the topological charges $S_1$ and $S_2$ have different signs to the one when sings are the same is described by the relation:

$$\beta = \frac{\lambda_1 - \gamma\lambda_2}{\lambda_1 + \gamma\lambda_2}. \quad (15)$$

For intensities equality ($\gamma = 1$) and wavelengths $\lambda_1 = 0.63\ \mu m$ and $\lambda_2 = 0.53\ \mu m$, $\beta = 0.07$. The value of the resulting Poynting vector at the superposition of two vortices with different charges is only 7% of the same value if the vortices interact with charges of the same sign.

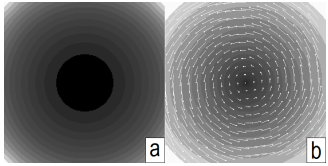

Fig. 2. The averaged Poynting vector for the resulting field, formed as a superposition of vortices with different frequencies and the same topological charges.
*a)* – the intensity; *b)* – the distribution of averaged Poynting vector characteristics.

Figure 2 shows the total intensity resulting from the superposition of vortices with different frequencies and the same topological charges and the corresponding behavior of component of the Poynting vector.

The total intensity resulting from the superposition of vortices with the different signs and the corresponding behavior of the averaged Poynting vector component practically does not differ in structure from the previous case. But the mean value of the component of the Poynting vector is much smaller.

It should be noted that with an appropriate choice of intensities of interacting vortices with different charges, it is possible to "quench" the transverse component of the Poynting vector. In this case, the exact zero of intensity remains in the center of the field, i.e. if such a beam is focused, then a vortex-free (absent orbital field moment) dark polychromatic optical trap is formed.

To verify the obtained theoretical estimations, we experimented with the following experimental arrangement (figure 3), which allows us to estimate the ratio of the orbital momenta, both produced by beams with a single radiation wavelength, and to evaluate the resulting orbital moment in a polychromatic beam.

Radiations with different frequencies were formed using a semiconductor laser (1) with frequency doubling ($\lambda_1 = 0.53\ \mu m$) and a He-Ne laser (11) ($\lambda_2 = 0.63\ \mu m$). For adjusting the intensity of the beams and to correcting the distribution parameters of this characteristic in each beam a composition of optical elements was used: two polarizers (intensity regulator) in which the orientation of the output polarizers was coordinated and formed collinear directions of oscillation of the field vectors for both frequencies and telescopic systems with pinholes specifically matched sizes.

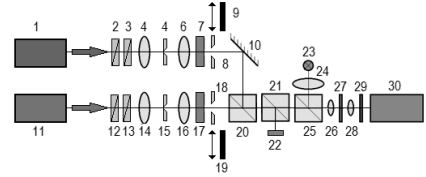

Fig. 3. Optical arrangement for checking theoretical estimation on the formation of the resulting Poynting vector by superposition of different frequencies waves

After passing through the correcting system, the radiation in both cases entered the vortex computer-generated holograms. Herewith, the corresponding orientation of the holograms made it possible to form a vortex with a certain sign of the topological charge. Then, the transverse dimensions of the beams were limited by apertures (8) and (18). Moreover, the sizes of the apertures were selected so that after focusing the beams in the rear focal plane of the micro-lens (26), they formed optical traps of the same size.

The alignment of the traps and their identical spatial localization was ensured by the corresponding adjustment of the mirror (10) and the beam splitter (20). The main radiation forming the traps was suppressed using the filter (29). The result of the microparticles capture was observed using the (28,30) system. The manipulation area was highlighted using the (23,24,25) system. The intensity level of the beams was monitored by a photo detector (20). Any of the branches of the circuit ("red" or "green") could be easily blocked using screens (9) and (19), which allowed exposure to the particle by radiation of one of the wavelengths or by a polychromatic wave with two wavelengths.

Obviously, the main reason for the mismatch of the experimental situation under the action of a polychromatic optical trap on a particle is the mismatch between the localizations of the red and green traps.

To analyze how the shift of one trap relative to another will affect the final distribution of the transverse component of the Poynting vector, we carried out the corresponding computer simulation.

Figure 4 show the changes in the intensities distribution and the distribution of the characteristics of the Poynting vector transverse component of the resultant field for cases when the charges of partial beams have the same and different signs.

The beam intensities were assumed to be the same. Beam widths were also assumed identical. The shift of the partial vortices was estimated in parts of the beam width.

As to the intensity, the changes in its distribution during the shift of partial vortices do not depend on the ratio of the signs of topological charges. First, a single minimum is blurred. In the next stage, two separate minima arise. It should be noted that the intensity at the minima is not a priori equal to zero and equal to the intensity that an alternative beam creates at this point.

As for the behavior of the Poynting vector, here the situation is completely different. For the superposition of beams with the same sign of the topological charge, even with significant shifts, the Poynting vector circulation will be preserved. In this case, a slight decrease in the transverse component value is observed.

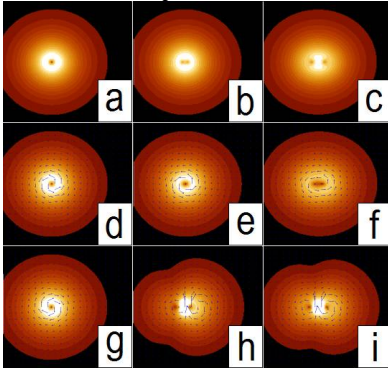

Fig. 4. Changes of the intensity (*a-c*) and characteristics of the Poynting vector (*d-i*) for the resulting field due to the shift of the partial vortices; (*d-f*) – the vortices charges are the same; (*g-h*) – the vortices charges are different. (*a, d, g*) – correspond to the case, when shift between vortices is absent; (*b, e, h*) – magnitude of shift is equal to the 0.04 of beam width; (*c, f, i*) – magnitude of shift is equal to the 0.1 of beam width.

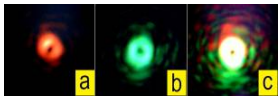

Fig. 5. The structure of the formed traps for different wavelengths (*a, b*) and the multi-frequency trap (*c*).

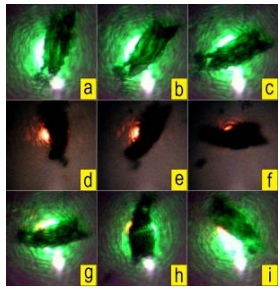

Fig. 6. Rotation of a "giant" particle (soot) captured by a dark vortex trap, which was formed by radiation with a single wavelength: green (*a-c*), red (*d-f*) and polychromatic trap (*g-i*).

Herewith a superposition of vortices of different signs, even small shifts (of the order of 4% of the beam width) leads to the radical changes in the distribution of the characteristics of the Poynting vector and increasing the total value of the component. Therefore, for experimental verification, we used a superposition of vortices with topological charges of the same sign.

Figure 5 shows the structure of the red, green, and multi-frequency traps. The following figures show us the behavior of a giant particle (soot) captured by a dark vortex trap, which can be formed by radiation with a single wavelength or polychromatic radiation. Figure 6 presents the behavior of a "giant" particle (soot) captured by a dark vortex trap, which was formed by radiation with a single wavelength and polychromatic trap.

It should be noted that the time intervals corresponding to horizontal changes in the figures are the same in all figures. Then:

1. The rotation speeds of the captured particle by traps formed by radiation with different wavelengths are approximately the same.

2. At the same time, the speed of rotation of a particle under the action of a polychromatic trap is much larger and approximately doubles the speed of particle rotation due to the action of traps formed by a single wavelength. This fact confirms the validity of the theoretical estimates made in the work.

## Conclusions

1. In a superposition of waves with different frequencies, the instantaneous components of the Poynting vector contain intermodulation components. Such components can have a significant effect on the formation of energy flows, when averaging is carried out over a time shorter than the beat time.

2. When averaging of the instantaneous Poynting vector over a longer beat-time, the components of the vector are a simple sum of elementary components associated with spectral components, regardless of whether a discrete or continuous set of waves takes part in the formation of the resulting field.

3. The angular momentum of such a wave can also be considered as an incoherent sum of elementary angular momenta of the field corresponding to each spectral component.

4. The obtained experimental results unambiguously confirm the given theoretical estimates.

**Disclosures**. The authors declare no conflicts of interest.